\documentclass[11pt]{article} 

\usepackage{amssymb}
\usepackage{amsmath}
\usepackage{mathdots}
\usepackage[dvips]{graphicx} 

\begin{document}
\sf



\noindent

\noindent

\noindent


\noindent


\noindent

\begin{center}

\center{\textbf{Pointlike electric charge in gravitational field theory}}

\noindent

\center{H. Dekker}

\noindent

\center{Delta Institute for Theoretical Physics, University of Amsterdam, Science Park,\\ The Netherlands}\\ \center{Private Institute for Advanced Study, R\'{e}sidence Le Jardin, Amsterdam,\\ The Netherlands}

\noindent

\center{hadekk944@gmail.com}

\end{center}

\noindent \textbf{Abstract}
\\

\noindent The existence of charged elementary 'point particles' still is a basically unsolved puzzle in theoretical physics. The present work takes a fresh look at the problem by including gravity---without resorting to string theory. Using Einstein's equations for the gravitational fields in a general static isotropic metric with the full energy-momentum tensor (for the charged material mass and the electromagnetic fields) as the source term, a novel exact solution with a well-defined characteristic radius emerges where mass and charge accumulate: $r_{\rm c}{=}\sqrt{r_{\rm e}r_o/2}$---with $r_{\rm e}{=}Q^2\!/4\pi\epsilon_omc^2$ being the 'classical' radius associated with the total charge $Q$ and where $r_o{=}2mG/c^2$ is the Schwarzschild radius belonging to the observable mass $m$ (for the electron one has $r_{\rm e}{\approx}10^{-15}$m and $r_o{\approx}\,10^{-57}$m). The resulting 'Einstein-Maxwell' gravitational electron radius can also be written as $r_{\rm c}{=}\ell_{\rm P}\sqrt{\alpha_{\rm e}}$, where $\ell_{\rm P}{=}\sqrt{\hbar G/c^3}{\approx}10^{-35}$m is the fundamental Planck length and $\alpha_{\rm e}{=}e^2\!/4\pi\epsilon_o\hbar c{\approx}1/137$ the fine-structure constant, which yields $r_{\rm c}^{\rm electron}{=}1.38063{\times}10^{-36}$m.

\noindent\\ \textbf{Keywords:} Point charge, self-gravitation, novel exact solution, Einstein-Maxwell electron radius, renormalization

\noindent\\ \textbf{PACS numbers:} 04.20.Cv, 04.20.Jb, 04.20.-q, 04.40.Nr, 04.40.-b



\noindent\\ Language: English

\noindent\\ Date of Submission: August 30, 2018





\noindent\\ Journal: Journal of Advances in Physics

\noindent\\ Website: https://cirworld.com

\noindent\\ \includegraphics*[width=0.92in, height=0.32in, keepaspectratio=false]{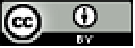} \nopagebreak This work is licensed under a Creative Commons Attribution 4.0 International License.

\noindent

\noindent

\noindent

\noindent

\noindent\newpage
\noindent \textbf{I. Introduction}\\


Modern theoretical physics is essentially based on the existence of a finite set of elementary 'point particles'---leptons and quarks---and their electromagnetic, gravitational, and weak or strong interactions (see, e.g., Refs.~[1--5]). Apart from the neutrino's, all fundamental particles carry an electric charge. However, the very concept of a stable 'point charge'---such as the electron---is an old and as yet basically unsolved problem. Namely, why should it be possible to accumulate a finite amount of electric charge in an infinitely small volume? What internal force does the work against the repulsive self-interaction? In fact, without such a force the charged particle should immediately explode.

Historically, the problems with a point charge were already recognized in classical physics (see, e.g., Refs.~[6--12]). For instance, in Poincar\'{e}'s 'electron model'~\cite{poincare} the electric force on the charged sphere was counteracted by an elastic force of unspecified, non-electromagnetic and non-gravitational nature in order to define a total energy-momentum tensor $T_{\alpha\beta}$ satisfying the condition $\partial T_{\alpha\beta}/\partial x_\beta{=}0$ characteristic of a closed system (see, e.g., Ref.~\cite{moller},~Ch.~7.3). In fact, even more than a decade after the advent of general relativity, during a visit to Leiden University in 1925, Einstein asked Lorentz' opinion on a purely electromagnetic model for the electron---i.e., without gravitational forces. Lorentz, however, rejected the idea (Ref.~\cite{kox}, Letter~398).

In any case, to quote from Feynman's Lectures (Ref.~\cite{feynman} Vol.~I, p.~32-4): ``The classical electron radius $r_{\rm e}{=}e^2 \!/4\pi \epsilon_o mc^2$ $\approx10^{-15}$m no longer has the significance that we believe that the electron really has such a radius''. More recently, based on state-of-the-art precision measurements of the electron's gyromagnetic  $g$-factor (and using a simple 'electron model' due to Brodsky and Drell \cite{br&dr}), 1989 Nobel laureate H.G.~Dehmelt  has pointed out that "Today everybody 'knows' the electron is a Dirac point particle with radius $r_*{=}0$ and $g{=}2$. But is it? The value $r_*{\approx}10^{-22}$ m given here constitutes an important new upper limit. [..] Thus, the electron may have size." (Ref.~\cite{dehmelt}; see also, e.g., Refs.~\cite{odom,gabrielse}).

Nevertheless, in non-gravitational quantum theory the electron can be treat\-ed successfully as a structureless point particle---at least, if the problem of its infinite self-energy is being 'swept under the carpet'. Indeed, as is well-known from the very beginning of quantum electrodynamics (see, e.g, Ref.~\cite{kramers}), this 'success' is only possible at a price. Namely, handling the infinite self-energy of a point charge requires an infinite mass renormalization to yield results in terms of the observed mass $m$ (see, e.g., Refs.~[3,4,18--23]). Unfortunately, this fundamentally hampers the unification with quantum gravity---as the latter has been found to be non-renormalizable.

For instance, to quote from Ref.~\cite{zinn},~p.~568: ``The definition of quantum gravity beyond the formal level leads to a number of unsolved problems. [...] Even pure quantum gravity is non-renormalizable in four dimensions. In fact, it is commonly believed that the theory remains non-renormalizable, a property which would indicate the breakdown of local quantum theory at Planck's length scale $\ell_{\rm P}{=}\sqrt{\hbar G/c^3}{\approx}10^{-35}$m.'' It is further worth noticing that modern string theory has been proposed---and is still under development---inter alia to cope with the problem of point-like particles, replacing them by tiny one-dimensional loops of Planck size (see, e.g., Refs.~\cite{becker,wiki1}).

The present work takes a fresh look at the problem by including gravity---as yet without resorting to string theory, ignoring the weak interaction and without a priori assuming the mass-charge density to be rigorously zero outside some $r_{\rm m}$ (as in the usual Reissner-Nordstr\"{o}m analysis; see, e.g., Refs.~\cite{moller,reissner,nordstrom} and Appendix~A). Namely, the enormous amount of electrostatic energy compressed into an infinitely small volume must---according to Einstein's general theory of relativity---give rise to huge local gravitational effects. Therefore, in this article the gravitational field equations for the Ricci tensor $R_{\mu\nu}$ are studied for a classical self-gravitating charged mass, i.e., with the full energy-momentum tensor $T^{\mu\nu}$ for the material mass and the electromagnetic fields as the source of gravitational energy---in a static isotropic metric (Sec.~II and~III). The analysis involves an 'electrostatic equilibrium' condition (Sec.~IV), and rigorously yields a well-defined novel mass-charge distribution with (in fact, for all charged leptons) a characteristic size $r_{\rm c}{\approx}10^{-36}$m (Sec.~V and~VI).

\noindent\\
\textbf{II. The gravitational field equations}\\
\renewcommand{\theequation}{\arabic{section}.\arabic{equation}}
\setcounter{section}{2}
\setcounter{equation}{0}


Throughout the present article, it is chosen to keep both the speed of light $c$ and the gravitational constant $G$ explicitly in the formula---rather than using 'geometrized units' where $c{=}G{=}1$. The notation closely follows that of Weinberg's book \cite{weinberg}. The field equations of general relativity may then be written as
\begin{equation}
R_{\mu\nu} = -\frac{8\pi}{c^4}G_{\mu\nu},
\label{2.1}
\end{equation}
with
\begin{equation}
G_{\mu\nu} = G \left( T_{\mu\nu}-\frac{1}{2}g_{\mu\nu}{T^\lambda}_\lambda \right),
\label{2.2}
\end{equation}
where $R_{\mu\nu}$ is the Ricci tensor, $T_{\!\mu\nu}{=}g_{\mu\lambda}g_{\nu\kappa}T^{\lambda\kappa}$ the covariant energy-momentum tensor, $g_{\mu\nu}$ the metric tensor, and ${T^\lambda}_\lambda{=}g^{\mu\nu}T_{\mu\nu}$ (with $g^{\mu\nu} g_{\nu\lambda}{=} {\delta^\mu}_\lambda$). In mixed components (see, e.g., Sec.~IV), Eq.~(\ref{2.1}) becomes ${R^\mu}_\nu{=}{-}(8\pi/c^4){G^\mu}_\nu$ with ${R^\mu}_\nu{=}$ $g^{\mu\lambda}R_{\lambda\nu}$ and ${G^\mu}_\nu{=}G({T^\mu}_\nu{-}\frac{1}{2}{\delta^\mu}_\nu{T^\lambda}_\lambda)$. The covariant tensor $g_{\mu\nu}$ defines the Riemannian space-time geometry by means of the proper time $d\tau$, such that
\begin{equation}
d\tau^2 = -c^{-2}g_{\mu\nu} dx^\mu dx^\nu,
\label{2.3}
\end{equation}
where, in this paper, time will be labelled by $\mu{=}0$.

For the problem of a charged mass, the energy-momentum tensor consists of two contributions, viz., for the---electrically charged---matter and the electromagnetic field itself. In the standard 'ideal fluid' form (without internal pressure), the energy-momentum tensor for the mass reads
\begin{equation}
T^{\mu\nu}_{\rm m} = \rho_{\rm m} \frac{dx^\mu}{d\tau} \frac{dx^\nu}{d\tau},
\label{2.4}
\end{equation}
where $\rho_{\rm m}{=}{-}c^{-2}T_{{\rm m}\,\lambda}^\lambda$ is the proper mass density. In the Dirac representation one has
\begin{equation}
\rho_{\rm m} = g^{-1/2}\sum_n m_n \int \delta^4[x-x_n(\tau)] d\tau,
\label{2.4a}
\end{equation}
where $g{=}\|{-}g_{\mu\nu}\|$ is the determinant of the metric tensor, so that the mass $M{=}$ $\sum_n m_n$ is given by
\begin{equation}
M = \int \gamma^{1/2} \rho_{\rm m}\, d^3x,
\label{2.5}
\end{equation}
where $\gamma{=}{-}g/g_{oo}$ is the determinant of the three-dimen{\-}sional metric tensor $\gamma_{ij}{=}$ $(g_{ij}{-}g_{oi}g_{oj}/g_{oo})$---see, e.g., Refs.~\cite{moller,landau&lifshitz2}.

The system is closed (so that $T^{\mu\nu}_{{\rm total};\nu}{=}0$) by including the electromagnetic field energy-momentum, given by
\begin{equation}
T^{\mu\nu}_{\rm em} = \epsilon_o \left( {F^\mu}_\lambda F^{\nu\lambda} -\frac{1}{4} g^{\mu\nu} F_{\lambda\kappa}F^{\lambda\kappa} \right),
\label{2.10}
\end{equation}
where the antisymmetric electromagnetic field tensor $F^{\mu\nu}$ is defined by $F^{oi}{=}E_i$ for the electric field and $F^{ij}{/}c$ ${=}\varepsilon_{ijk}B_k$ for the magnetic field, $\varepsilon_{ijk}$ representing the usual three-dimensional Levi-Civita symbol and $\epsilon_o$ being the vacuum permittivity. It is useful to note that, since $g_{\mu\nu}{F^\mu}_\lambda{=}F_{\nu\lambda}$ and $g_{\mu\nu}g^{\mu\nu}{=}4$, Eq.~(\ref{2.10}) implies that $T^\lambda_{{\rm em}\lambda}{=}g_{\mu\nu}T^{\mu\nu}_{\rm em}{=}0$.

The electromagnetic fields satisfy Maxwell's equations
\begin{equation}
\frac{\partial}{\partial x^\nu}\left( g^{1/2}F^{\mu\nu} \right) = g^{1/2} J^\mu_{\rm em}/\epsilon_o,
\label{2.8}
\end{equation}
where $J_{\rm em}^\mu{=}c^{-1}\rho_{\rm e} dx^\mu{/}d\tau$ is the current four-vector---with $\rho_{\rm e}$ being the proper charge density, which in the Dirac representation reads
\begin{equation}
\rho_{\rm e} = g^{-1/2}\sum_n e_n \int \delta^4[x-x_n(\tau)] d\tau.
\label{2.9a}
\end{equation}
Since the current satisfies the conservation law $J^\mu_{{\rm em};\mu}{=}0$, the charge $Q{=}\sum_n e_n$ is conserved and given by
\begin{equation}
Q = \int \gamma^{1/2} \rho_{\rm e}\, d^3x.
\label{2.9}
\end{equation}
Note that both $\rho_{\rm e}$ and $\rho_{\rm m}$ transform like a scalar.

\noindent\\
\textbf{III. The static isotropic case}\\
\setcounter{section}{3}
\setcounter{equation}{0}

The static isotropic metric may be written in the 'standard' form in spherical coordinates $r,\theta,\varphi$, so that the only nonvanishing components of the metric tensor are
\begin{equation}
g_{oo}\!=\!-B(r),\,\, g_{rr}\!=\!A(r),\,\, g_{\theta\theta}\!=\!r^2,\,\, g_{\varphi\varphi}\!=\!r^2{\rm sin}^2\theta,
\label{3.1}
\end{equation}
and $g{=}A(r)B(r)r^4{\rm sin}^2\theta$. The only nonzero components of the Ricci tensor are (with $A^\prime{=}dA/dr$, etc.)
\begin{eqnarray}
R_{oo} &=& -\frac{B^{\prime\prime}}{2A} +\frac{B^\prime}{4A} \left( \frac{A^\prime}{A}+\frac{B^\prime}{B} \right) -\frac{B^\prime}{rA}, \label{3.2} \\
R_{rr} &=& \frac{B^{\prime\prime}}{2B} -\frac{B^\prime}{4B} \left( \frac{A^\prime}{A}+\frac{B^\prime}{B} \right) -\frac{A^\prime}{rA}, \label{3.3} \\
R_{\theta\theta} &=& -1 -\frac{r}{2A} \left( \frac{A^\prime}{A}-\frac{B^\prime}{B} \right) +\frac{1}{A},
\label{3.4}
\end{eqnarray}
while $R_{\varphi\varphi}{=}R_{\theta\theta}{\rm sin}^2\theta$ and the only nonzero component of the material energy-momentum tensor now reads
\begin{equation}
T^{oo}_{\rm m} = \rho_{\rm m}c^2/B,
\label{3.5}
\end{equation}
which readily yields
\begin{eqnarray}
G^{\rm m}_{oo}=\frac{1}{2} G B\rho_{\rm m}c^2&,& G^{\rm m}_{rr}=\frac{1}{2} G A\rho_{\rm m}c^2, \nonumber \\
G^{\rm m}_{\theta\theta}&=&\frac{1}{2} G r^2 \rho_{\rm m}c^2,
\label{3.6}
\end{eqnarray}
while $G^{\rm m}_{\varphi\varphi}{=}G^{\rm m}_{\theta\theta}{\rm sin}^2\theta$.

For the metric (\ref{3.1}), Eq.~(\ref{2.8}) leads to the Poisson equation
\begin{equation}
\frac{d}{dr}\left( r^2 \sqrt{AB} E_r \right) = r^2 \sqrt{A} \,\frac{\rho_{\rm e}}{\epsilon_o}
\label{3.10}
\end{equation}
for the only nonzero component $E_r{=}F^{or}$ of the electric field, while for the contributions from Eq.~(\ref{2.10}) one obtains
\begin{eqnarray}
G^{\rm em}_{oo}=\frac{\epsilon_o}{2} GAB^2 E_r^2&,&  G^{\rm em}_{rr}=-\frac{\epsilon_o}{2} GA^2B E_r^2, \nonumber \\
G^{\rm em}_{\theta\theta}&=&\frac{\epsilon_o}{2} GABr^2 E_r^2,
\label{3.12}
\end{eqnarray}
while $G^{\rm em}_{\varphi\varphi}{=}G^{\rm em}_{\theta\theta}{\rm sin}^2\theta$. Note that $G_{{\rm em}\lambda}^\lambda{=}g^{\mu\nu}G^{\rm em}_{\mu\nu}{=}0$, as it should be.

\noindent\\
\textbf{IV. The equilibrium condition}\\
\setcounter{section}{4}
\setcounter{equation}{0}

Einstein's gravitational field equations (\ref{2.1}) for the static isotropic mass-charge system may thus be written as
\begin{eqnarray}
\frac{R_{oo}}{B} &=& -\frac{4\pi}{c^2} G\rho_{\rm m} -\frac{4\pi\epsilon_o}{c^4} GAB E_r^2, \label{4.1} \\
\frac{R_{rr}}{A} &=& -\frac{4\pi}{c^2} G\rho_{\rm m} +\frac{4\pi\epsilon_o}{c^4} GAB E_r^2, \label{4.2} \\
\frac{R_{\theta\theta}}{r^2} &=& -\frac{4\pi}{c^2} G\rho_{\rm m} -\frac{4\pi\epsilon_o}{c^4} GAB E_r^2.
\label{4.3}
\end{eqnarray}
Note that $R_{oo}/B{=}{-}{R^o}_o$, $R_{rr}/A{=}{R^r}_r$, and $R_{\theta\theta}/r^2{=}{R^\theta}_\theta$.

To find the equilibrium equation, first consider $R_{rr}/A{+}R_{oo}/B$. This leads to
\begin{equation}
\frac{1}{A} \left( \frac{A^\prime}{A}+\frac{B^\prime}{B} \right) = \frac{8\pi}{c^2} Gr\rho_{\rm m}.
\label{4.4}
\end{equation}
Using the $R_{\theta\theta}$-equation (\ref{4.3}) to eliminate $A^\prime$, one gets
\begin{equation}
\frac{r}{A} \frac{B^\prime}{B} +\frac{1}{A} = 1-\frac{4\pi\epsilon_o}{c^4} GAB r^2 E_r^2.
\label{4.5}
\end{equation}
Now differentiating Eq.~(\ref{4.5}) with respect to $r$ and using $R_{rr}/A{-}R_{oo}/B$ to eliminate $B^{\prime\prime}$, one obtains
\begin{eqnarray}
\hspace{-6mm}\frac{rB^\prime}{2AB} \left( \frac{A^\prime}{A}+\frac{B^\prime}{B} \right) =& & \frac{8\pi\epsilon_o}{c^4} GAB r E_r^2 \nonumber \\
&+& \frac{4\pi\epsilon_o}{c^4} G \frac{d}{dr}\left( AB r^2 E_r^2 \right).
\label{4.6}
\end{eqnarray}
Once more invoking Eq.~(\ref{4.4}), the result reads
\begin{equation}
\frac{B^\prime}{B} = \frac{2AB r E_r^2 +d(AB r^2 E_r^2)/dr}{\rho_{\rm m} \,c^2 r^2 /\epsilon_o },
\label{4.7}
\end{equation}
which---putting ${\cal E}{=}r^2\sqrt{AB}E_r$ and noticing the identity $d({\cal E}^2\!/r^2){/}dr{=}{-}2{\cal E}^2\!/r^3{+}$ $2({\cal E}\!/r^2)\,d{\cal E}\!/dr$ ---by invoking the Poisson equation (\ref{3.10}) can be rewritten as
\begin{equation}
\frac{B^\prime}{B} = \frac{2 A \sqrt{B} E_r \rho_{\rm e}}{\rho_{\rm m} \,c^2},
\label{4.8}
\end{equation}
which represents the balancing of the repulsive electrostatic self-force and the attractive gravitational self-force. It is the electrostatic counterpart of the usual 'hydrostatic equilibrium' condition for ideal fluids. In fact, by virtue of the Bianchi identities, Eq.~(\ref{4.8}) is a direct consequence of the conservation law $T_{{\rm total},\nu}^{r\nu}{=}0$. Namely, one gets $T_{{\rm m};\nu}^{r\nu}{=}\Gamma_{oo}^rT_{\rm m}^{oo}{=}\Gamma_{oo}^r \rho_{\rm m}\, c^2/B$ with $\Gamma_{oo}^r{=}B^\prime/2A$, while by means of the Poisson equation (\ref{3.10}) one obtains $T_{{\rm em};\nu}^{r\nu}{=}\sqrt{B}E_r\rho_{\rm e}$.

For a structureless charged mass the intrinsic charge-to-mass ratio $e_n/m_n$ should be an $n$-independent constant, i.e., $e_n{=}{\sf k}m_n$, which according to Eqs.~(\ref{2.4a}) and (\ref{2.9a}) implies the equation of state $\rho_{\rm e}{=}{\sf k}\rho_{\rm m}$ for the proper density. Without loss of generality, one may put ${\sf k}{=}Q/M$ so that---by virtue of Eq.~(\ref{2.9}) for $Q$---the as yet undetermined mass $M$ satisfies Eq.~(\ref{2.5}). Hence, the equilibrium equation (\ref{4.8}) may be rewritten as
\begin{equation}
AE_r=\frac{\mu}{2}\frac{B^\prime}{B^{3/2}},
\label{4.10}
\end{equation}
with $\mu{=}Mc^2/Q$. Now using the Newtonian limit $B{=}$ $1{-}2mG/rc^2$ and the Poisson limit $AE_r{=}$ $Q/4\pi \epsilon_o r^2$  for $r{\rightarrow}\infty$, one obtains
\begin{equation}
m = \frac{Q^2}{4\pi\epsilon_o G M}.
\label{4.13}
\end{equation}
By Eq.~(\ref{4.10}) the problem of the charged mass is reduced to finding the temporal metric function $B(r)$.

\noindent\\
\textbf{V. The temporal metric function}\\
\setcounter{section}{5}
\setcounter{equation}{0}

Consider Eq.~(\ref{4.5}) for $A$, and using Eq.~(\ref{4.10}) for $E_r$ write it as
\begin{equation}
A = 1 +\beta +{\sf M}\beta^2,
\label{5.1}
\end{equation}
where $\beta{=}rB^{\prime}/B$ and ${\sf M}{=}\pi\epsilon_o G \mu^2/c^4$. Now again take $R_{rr}/A{-}R_{oo}/B$, collect the $A^\prime$ terms and once more use Eq.~(\ref{4.10}) for $E_r$. After a somewhat laborious but otherwise elementary calculation this leads to
\begin{equation}
\frac{rB^{\prime\prime}}{B} +\left( 1 -\frac{1}{2}\beta \right) \frac{B^\prime}{B} -\left( 1 +\frac{1}{2}\beta \right) \frac{A^\prime}{A} = 2{\sf M}\beta \frac{B^\prime}{B}.
\label{5.2}
\end{equation}
Substituting $A$ and $A^\prime$ from Eq.~(\ref{5.1}), and combining similar terms---some of which add up to zero---Eq.~(\ref{5.2}) is found to factorize such that for ${\sf M}{=}\frac{1}{4}$ it becomes a trivial zero identity (see Appendix~A and, e.g., Ref.~\cite{arnowitt}) while for ${\sf M}{\ne}\frac{1}{4}$ it either leads to the trivial solution $B^\prime{=}0$ or to a nontrivial metric function $B(r)$ satisfying
\begin{eqnarray}
\frac{1}{2}rB^{\prime\prime}
+\left( 1+\frac{1}{2}{\sf M}\beta^2 \right) B^\prime =0,
\label{5.3}
\end{eqnarray}
which is akin to the prototype equation $d\beta/dt{=}\beta^3$ for the temporal evolution of so-called 'finite-time blow-up' processes in, e.g., chemistry and hydrodynamic turbulence (see, e.g., Ref.~\cite{hadekk3}, p.~353).

Noticing that $B\beta^\prime{=}rB^{\prime\prime}{+}(1{-}\beta)B^\prime$, using Eq.~(\ref{5.3}) for $B^{\prime\prime}$ and defining the auxiliary variable $\zeta(r){=}1/\beta$, one thus rigorously obtains
\begin{equation}
r \zeta \zeta^\prime = {\sf M} +\zeta +\zeta^2.
\label{6.2}
\end{equation}
The exact solution of Eq.~(\ref{6.2}) is given by
\begin{equation}
\frac{1}{2} {\rm ln}\left( {\sf M}\! +\!\zeta\! +\!\zeta^2 \right)\!+\!\frac{1}{2\sqrt{{\sf M}{-}\frac{1}{4}}}\, {\rm arccot} \frac{\zeta{+}\frac{1}{2}}{\sqrt{{\sf M}{-}\frac{1}{4}}} \!=\!{\rm ln}\left(\! \frac{r}{r_o} \!\right),
\label{6.3}
\end{equation}
where the integration constant has been set equal to the Schwarzschild radius $r_o{=}2mG/c^2$ belonging to the observed mass m in order to satisfy the Newtonian limit $\zeta{\approx}r/r_o$ for large values of $r$.

For $r{\rightarrow}\infty$, one has $\zeta{\approx}r/r_o{\rightarrow}\infty$ as well. On the other hand, $\zeta{\rightarrow}0$ for $r{\rightarrow}r_{\rm c}$. Namely, expanding Eq.~(\ref{6.3}) in powers of $\zeta$ yields $\zeta^2/2{\sf M}{\approx}{\rm ln}(r/r_{\rm c})$, with
\begin{equation}
r_{\rm c} = r_o \sqrt{\sf M}\, {\rm exp}\!\left( \frac{1}{2\sqrt{{\sf M}{-}\frac{1}{4}}} \,{\rm arccot} \frac{1}{2\sqrt{{\sf M}{-}\frac{1}{4}}} \right).
\label{6.4}
\end{equation}
Using $\mu{=}Mc^2/e$ [as given below Eq.~(\ref{4.10})] in the definition of {\sf M} [as given below Eq.~(\ref{5.1})] and invoking Eq.~(\ref{4.13}) for $M$, one obtains ${\sf M}{=}M/4m$. Once more using (\ref{4.13}), this becomes ${\sf M}{=}Q^2/16\pi \epsilon_o G m^2$ which leads to ${\sf M}{=}r_{\rm e}/2r_o$---with $r_{\rm e}{=}Q^2\!/4\pi\epsilon_omc^2$ being the 'classical' radius belonging to the charge $Q$ and where $r_o{=}2mG/c^2$ is the Schwarzschild radius belonging to the mass $m$---so that $r_{\rm c}{\approx}\sqrt{r_or_{\rm e}/2}$, which explicitly amounts to
\begin{equation}
r_{\rm c} = |Q|\sqrt{\frac{G}{4\pi \epsilon_o c^4}}\, ,
\label{6.5}
\end{equation}
where the exponential factor from Eq.~(\ref{6.4}) has been omitted as it is always of the order of unity. Actually, with $r_{\rm e}{\approx}10^{-15}$m and $r_o{\approx}10^{-57}$m for the electron, one has ${\sf M}{=}r_{\rm e}/2r_o{\approx}10^{42}$  so that the exponential correction tends to ${\rm exp}(\pi/4\sqrt{\sf M})$ ${\approx}{1}{+}{\cal O}\,(10^{-21})$. Finally, one obtains $r_{\rm c}^{\rm electron}{\approx}10^{-36}$m---which may be called the 'Einstein-Maxwell' gravitational electron radius.



\noindent\\
\textbf{VI. Results}\\
\setcounter{section}{6}
\setcounter{equation}{0}

The temporal metric function now follows from $dB/d\zeta{=}B/r\zeta\zeta^\prime$. Using Eq.~(\ref{6.2}) for $\zeta^\prime$, this yields
\begin{equation}
B = {\rm exp}\!\left(\! -\frac{1}{\sqrt{{\sf M}{-}\frac{1}{4}}}\, {\rm arccot} \frac{\zeta{ +}\frac{1}{2}}{\sqrt{{\sf M}{-}\frac{1}{4}}} \right),
\label{6.6}
\end{equation}
where $\zeta(r/r_{\rm c})$ follows from Eqs.~(\ref{6.3}) and (\ref{6.4}), while $B(r{\leq}r_{\rm c}){=}B_{\rm c}$. Notice that for finite values of $r/r_{\rm c}$ and ${\sf M}{\rightarrow}\infty$ one has $\zeta/\sqrt{\sf M}{=}\sqrt{(r/r_{\rm c})^2{-}1}$, so that the function ${\cal B}(r){=}(B{-}B_{\rm c})/(1{-}B_{\rm c})$ becomes ${\cal B}{=}(2/\pi){\rm arccos}(r_{\rm c}/r)$. Fig.~1 shows the exact solution $B(r)$ of Eq.~(\ref{5.3}) for a few values of {\sf M}. Further, using Eq.~(\ref{6.3}) for $\zeta$, Eq.~(\ref{6.6}) can also be written as $B(r){=}(r_o/r)^2({\sf M}{+}\zeta{+}\zeta^2)$. Hence, $B_{\rm c}{=}(r_o\sqrt{\sf M}/r_{\rm c})^2$ while a simple calculation using Eqs.~(\ref{6.2}) and (\ref{6.3}) yields $B^\prime(r{\rightarrow}r_{\rm c}){\propto}\, 1/\sqrt{r{-}r_{\rm c}}$.
\begin{figure}
\begin{center}
\includegraphics{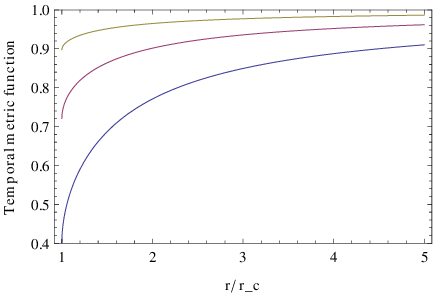}
\end{center}
\sf Figure 1. The temporal metric function $B(r)$ [for ${\sf M}{=}2$ (bottom line), ${\sf M}{=}20$ (middle), ${\sf M}{=}200$ (top)]. Shown is the numerical solution of the exact Eq.~(\ref{5.3}), as a function of the non-dimensional radial variable $r/r_{\rm c}$.
\end{figure}

Similarly, the radial metric function $A(r)$ follows from Eq.~(\ref{5.1}) as
\begin{equation}
\frac{1}{A} = \frac{\zeta^2}{{\sf M} +\zeta +\zeta^2},
\label{6.8}
\end{equation}
while $A(r{<}r_{\rm c}){=}1$. For finite $r/r_{\rm c}$ and ${\sf M}{\rightarrow}\infty$, one gets $1/A{=}1{-}(r_{\rm c}/r)^2$. Next, the radial electrostatic field is obtained from Eq.~(\ref{4.10}), which yields
\begin{equation}
E_r = \frac{\mu}{2r\sqrt{B}} \,\frac{\zeta}{{\sf M}+ \zeta +\zeta^2} \,\Theta \left(r-r_{\rm c}\right),
\label{6.9}
\end{equation}
where $\Theta(r)$ is the unit step function. Fig.~2 shows the exact solution as $E_r(r)/E_{\rm c}$, with $E_{\rm c}{=}\mu/2r_{\rm c}\sqrt{\sf M}$. For finite values of $r/r_{\rm c}$ and with ${\sf M}{\rightarrow}\infty$, this function becomes $E_r/E_{\rm c}{=}(r_{\rm c}/r)^2\sqrt{1{-}(r_{\rm c}/r)^2}$, which has its peak value $\frac{2}{9}\sqrt{3}{\approx}0.38$ at $r/r_{\rm c}{=}$ $\frac{1}{2}\sqrt{6}{\approx}1.22$.
\begin{figure}
\begin{center}
\includegraphics{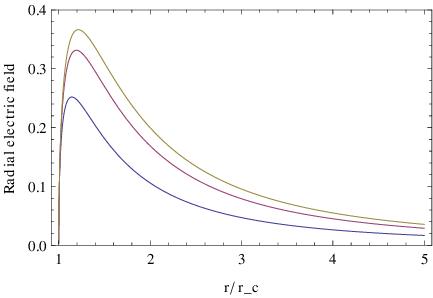}
\end{center}
\sf Figure 2. The radial electrostatic field $E_{\rm r}(r)$ [for ${\sf M}{=}2$ (bottom line), ${\sf M}{=}20$ (middle), ${\sf M}{=}200$ (top)]. Shown is $E_{\rm r}/E_{\rm c}$ [with $E_{\rm c}{=}\mu/2{\sf M}r_o$], using the numerical solution of the exact Eq.~(\ref{5.3}), as a function of the non-dimensional radial variable $r/r_{\rm c}$.
\end{figure}
One further gets
\begin{equation}
r^2 \sqrt{AB}E_r =  \frac{\mu}{2} \frac{r}{\sqrt{{\sf M}+ \zeta +\zeta^2}} \,\Theta \left(r-r_{\rm c}\right).
\label{6.10}
\end{equation}
Finally, the Poisson equation (\ref{3.10}) leads to
\begin{equation}
r^2 \sqrt{A} \rho_{\rm e} = \frac{\mu \epsilon_o r_{\rm c}}{2\sqrt{\sf M}} \,\delta (r-r_{\rm c}) -\frac{\mu \epsilon_o}{4\zeta} \frac{\Theta(r-r_{\rm c})}{\sqrt{{\sf M}+ \zeta +\zeta^2}},
\label{6.11}
\end{equation}
where $\delta(r)$ is the Dirac distribution and which rigorously satisfies Eq.~(\ref{2.9}) for the total charge $Q$---recalling that $\mu{=}Mc^2/Q$ [see its definition below Eq.~(\ref{4.10})] with $Mc^2{=}$ $Q^2/2\pi\epsilon_o r_o$ [from Eq.~(\ref{4.13}) and $r_o{=}2mG/c^2$] and using Eq.~(\ref{6.4}) for $r_{\rm c}$, so that
\begin{equation}
\frac{\mu \epsilon_o r_{\rm c}}{2\sqrt{\sf M}}=\frac{Q}{4\pi} {\rm exp}\!\left( \frac{1}{2\sqrt{{\sf M}{-}\frac{1}{4}}} \,{\rm arccot} \frac{1}{2\sqrt{{\sf M}{-}\frac{1}{4}}} \right).
\label{6.12}
\end{equation}
For finite values of $r/r_{\rm c}$ and ${\sf M}{\rightarrow}\infty$, the second part of the density (\ref{6.11}) now becomes $\rho_{\rm e}^{\rm cont}{=}$ ${-}[Q/8\pi r_{\rm c}^3\sqrt{\sf M}\,](r_{\rm c}/r)^4$. Clearly, with ${\sf M}{\approx}10^{42}$ [see below Eq.~(\ref{6.5})] charge and mass almost completely accumulate at the Einstein-Maxwell radius $r_{\rm c}^{\rm electron}$---the tail in Eq.~(\ref{6.11}) containing only some $10^{-21}Q$.

The singular part of the density follows from the Poisson equation rather than from the gravitational field equations per se. Namely, from Eq.~(\ref{6.11}) one has $\rho_{\rm e}^{\rm sing}{=}$ $\!(\mu\epsilon_o/2{\sf M}r_{\rm c})\,\zeta(r)\,\delta(r{-}r_{\rm c})$, so that by virtue of $\zeta(r_{\rm c}){=}0$ one has $T^{{\rm sing}\,oo}_{\rm m}{=}0$. Of course, the continuous part of the density $\rho_{\rm e}^{\rm cont}{=}{-}\mu\epsilon_o/4r^2\zeta^2A$ obeys the Einstein equations for $r{>}r_{\rm c}$ (for all values of {\sf M}). E.g., consider Eq.~(\ref{4.4}) and note that its right-hand side amounts to ${-}2{\sf M}/r\zeta^2A$, so that it remains to show that $r(A^\prime/A{+}B^\prime/B){=}{-}2{\sf M}/\zeta^2$---which is easily done since $rA^\prime/A{=}\zeta dA/d\zeta{=}{-}1/\zeta{-}$ $2{\sf M}/\zeta^2$ [by Eqs.~(\ref{6.2}) and (\ref{6.8})] and $rB^\prime/B{=}1/\zeta$ by definition.

\noindent\\
\textbf{VII. Final remarks}\\
\setcounter{section}{7}
\setcounter{equation}{0}

The mass-charge distribution (\ref{6.11}) is an exact particle-like solution of the classical Einstein-Maxwell equations (for all values of the parameter ${\sf M}{=}Q^2/16\pi \epsilon_o Gm^2$). It emerges from a rigorous balance between electrostatic self-repulsion and gravitational self-attraction for all values of the radius $r$ (see Sec.~IV), which is part of its novelty---see, e.g., Refs.~\cite{moller,reissner,nordstrom}. It is further worth noticing that it appears to be the only solution for a point-like charge which correctly satisfies the observable Newtonian and Poisson limits for large $r$ but does not necessarily explode (as is, for instance, the case for the usual Reissner-Nordstr\"{o}m 'superextremal' black hole electron; see, e.g., Ref.~\cite{reissner}).

For an elementary 'point charge'---like, e.g., the Dirac electron---the electrostatic self-interaction is the well-known source of an infinite self-energy (plaguing quantum field theory and its unification with general relativity), and it requires the inclusion of gravity to establish equilibrium at the new finite size $r_{\rm c}{=}\sqrt{r_{\rm e}r_o/2}{\approx}10^{-36}$m (where $r_{\rm e}{\approx}10^{-15}$m is the 'classical' electron radius and $r_o{\approx}10^{-57}$m is the electron Schwarzschild radius)---as shown at the end of Sec.~V. In fact, a system of this kind was first considered by Poincar{\'e} as a model for the electron, however, without specifying the nature of the model's counteracting 'elastic forces' (see, e.g., Refs.~\cite{poincare,moller}).

In the present work it is shown that these forces intrinsically arise from local gravitational effects, which become huge when compressing a finite charge into an infinitesimal volume and which give rise to the novel 'hidden' mass $M{=}e^2\!/4\pi\epsilon_omG$---which is essentially located at $r{=}r_{\rm c}$ while taking care of the observable Newtonian and Poisson limits for $r{\rightarrow}\infty$. For the electron---with the parameter ${\sf M}{=}M/4m{=}$ $r_{\rm e}/2r_o{\approx}10^{42}$ (while, e.g., for the heavier tauon ${\sf M}{\approx}10^{35}$)---mass and charge indeed almost completely accumulate at the radius $r_{\rm c}$ (see Sec.~VI).

By virtue of the relation (\ref{4.13}) between the point-like massive charge $Q/M$ and the observable charged mass $m/Q$, the characteristic radius $r_{\rm c}$ is mass independent (i.e., identical for all charged leptons). Namely, by Eqs.~(\ref{6.4}) and (\ref{6.5}) one has $r_{\rm c}{=}|e|\sqrt{G/4\pi\epsilon_o c^4}$---which is also worth noticing to follow at once from equating the electromagnetic mass $m_{\rm c}^{\rm charge}{=}e^2/8\pi\epsilon_o r_{\rm c}c^2$ at $r{=}r_{\rm c}$ to the Schwarzschild gravitational mass $m_{\rm c}^{\rm grav}{=}r_{\rm c}\,c^2/2G$ (so that for the parameter ${\sf M}$ one also has $1\!/\!\sqrt{\sf M}{=}m/m_{\rm c}^{\rm grav}$, with $m_{\rm c}^{\rm grav}{\approx}10^{-9}$kg). Hence, apart from involving the total charge, the novel 'Einstein-Maxwell' gravitational electron radius only depends on the fundamental constants $c$ and $G$. It is readily written as
\begin{equation}
r_{\rm c}^{\rm electron} = \ell_{\rm P} \sqrt{\alpha_{\rm e}},
\label{7.1}
\end{equation}
where $\ell_{\rm P}{=}\sqrt{\hbar G/c^3}{\approx}10^{-35}$m is the Planck length (see, e.g., Refs.~\cite{zinn,pais}) and $\alpha_{\rm e}{=}e^2\!/4\pi\epsilon_o\hbar c$ ${\approx}1/137$ is the fine-structure constant. The resulting numerical value (see, e.g., Ref.~\cite{wiki2}) is: $r_{\rm c}^{\rm electron}{=}1.38063{\times}10^{-36}$m.

Since the radius (\ref{7.1}) is about an order of magnitude smaller than Planck's length $\ell_{\rm P}$---which is the fundamental size of the tiny one-dimensional loops proposed in modern string theory in order to inter alia cope with the problem of point-like particles (see, e.g., Refs.~\cite{becker,wiki1})---the electron is indeed a point-like charge from the perspective of non-gravitational theory. However, while tiny it is large enough to produce only a relatively small quantum mechanical self-energy $\delta m$.

For instance, for the free electron (see, e.g., Ref.~\cite{sakurai}, p.~270, tentatively taking $\Lambda{\approx}1/r_{\rm c}$ for the ultraviolet wave number cutoff) one now qualitatively gets $\delta m/m\!\!\approx(3\alpha_{\rm e}/2\pi){\rm ln}(\ell_{\rm C}/r_{\rm c}){\approx}0.2$ (where $\ell_{\rm C}{=}\hbar/mc{\approx}10^{-13}$m is the Compton wavelength)---which removes the infinite mass renormalization from quantum electrodynamics (see also, e.g., Refs.~\cite{hadekk1,hadekk2}) and thus opens up new perspectives for unifying non-gravitational quantum field theory with non-renormalizable quantum gravity.

\appendix
\renewcommand{\theequation}{\Alph{section}.\arabic{equation}}
\setcounter{section}{1}
\setcounter{equation}{0}

\noindent\\
\textbf{Appendix}\\

\noindent
\textbf{The case ${\sf M}{=}\frac{1}{4}$}

For ${\sf M}{=}\frac{1}{4}$ the upshot from Eq.~({\ref{5.2}) yields a trivial zero identity for all metric functions $B(r)$. Hence, in that case the problem can be solved by any mass-charge distribution, the ensuing metric following from the Poisson equation (\ref{3.10}) with the equilibrium equation (\ref{4.10}) for $E_r$ and Eq.~(\ref{5.1}) for $A(r)$. Upon once integrating the Poisson equation, one gets
\begin{equation}
\frac{r}{r_o}\frac{\beta}{1{+}\frac{1}{2}\beta} = F_{\rm e}(r),
\label{A.1}
\end{equation}
with $F_{\rm e}{=}(4\pi/Q)\int_o^r r^2\sqrt{A}\rho_{\rm e}\,dr$---so that by Eq.~(\ref{2.9}) one has $F_{\rm e}(\infty){=}1$.\hspace{0.1mm} For instance, for a density with finite range $r_{\rm m}$ so that $\rho_{\rm e}(r{>}r_{\rm m}){\equiv}0$---such as, e.g., $4\pi r^2 \sqrt{A}\rho_{\rm e}{=}$ $Q\delta(r{-}r_{\rm m})$---one would have $\beta{=}1/(r/r_o{-}\frac{1}{2})$ for $r{>}r_{\rm m}$, and a metric singularity arises if $r_{\rm m}{<}\frac{1}{2}r_o$. \hspace{-2.3mm} With $\beta{=}rB^\prime\!/B$, this then leads to
\begin{equation}
B = \left( 1-\frac{1}{2}\frac{r_o}{r} \right)^2,
\label{A.2}
\end{equation}
which is worth noticing to correspond to a special case (viz., the 'extremal black hole') of the Reissner-Nordstr\"{o}m metric \cite{moller,reissner,nordstrom}.

A density with infinite range \hspace{-1mm} may be given by $F_{\rm e}{=}r/(r{+}\varepsilon r_o)$---i.e., $4\pi r^2 \sqrt{A}\rho_{\rm e}$ ${=}\varepsilon r_o/(r{+}\varepsilon r_o)^2$---which by Eq.~(\ref{A.1}) for ${\sf M}{=}\frac{1}{4}$ would imply $\beta{=}1/[r/r_o{+}(\varepsilon{-}\frac{1}{2})]$ and a singularity arises if $\varepsilon{<}\frac{1}{2}$. In this case the temporal metric function reads $B{=}[1{+}(\varepsilon {-}\frac{1}{2})r_o{/}r]^{{-}1/(\varepsilon{-}\frac{1}{2})}$, which for $\varepsilon{\rightarrow}0$ of course reduces to the result given in Eq.~(\ref{A.2}).

Finally, notice that by Eq.~(\ref{4.13}) for ${\sf M}{=}\frac{1}{4}$ the observable classical mass amounts to $m{=}$ \linebreak $|Q|/\sqrt{4\pi\epsilon_oG}$, so that in this case the corresponding Schwarzschild radius $r_o{=}2mG/c^2$ precisely equals the mass-independent 'Einstein-Maxwell' radius $r_{\rm c}{=}|Q|\sqrt{G/4\pi\epsilon_o c^4}{=}\ell_{\rm P}\sqrt{\alpha_{\rm e}}$ of Eq.~(\ref{7.1}). However, one now gets $m{\approx}10^{-9}$kg if $Q{=}e$---which is obviously far too heavy for the electron and is unlikely to be repaired by quantum effects (see, e.g., Sec.~7). To account for the correct order of magnitude of the electron mass $m{\approx}10^{-30}$kg one needs ${\sf M}{\gg}1$, as shown in the main text.

\renewcommand{\refname}{\normalsize {\boldmath $\sf References$}}

\normalsize

\end{document}